\documentclass[showpacs,twocolumn,amsmath,superscriptaddress]{revtex4-1}
\usepackage[dvipdfmx]{graphicx}
\usepackage{color,amsmath,lmodern,bm,array,braket}

\newcommand{\bk}{{\bm k}}
\newcommand{\bt}{{\bm t}}

\begin{document}


\title{Symmetry protected line nodes in non-symmorphic magnetic space groups: Applications to  UCoGe and UPd$_2$Al$_3$}%

\author{Takuya Nomoto}%
\email{nomoto.takuya@scphys.kyoto-u.ac.jp}
\affiliation{Department of Physics, Kyoto University, Kyoto, 606-8502, Japan}%
\author{Hiroaki Ikeda}%
\affiliation{Department of Physics, Ritsumeikan University, Kusatsu, 525-8577, Japan}%

\date{\today}%

\begin{abstract}
We present the group-thoretical classification of gap functions in superconductors coexisting with some magnetic order in non-symmorphic magnetic space groups. Based on the weak-coupling BCS theory, we show that UCoGe-type ferromagnetic superconductors must have horizontal line nodes on either $k_z=0$ or $\pm\pi/c$ plane. Moreover, it is likely that additional Weyl point nodes exist at the axial point. On the other hand, in UPd$_2$Al$_3$-type antiferromagnetic superconductors, gap functions with $A_g$ symmetry possess horizontal line nodes in antiferromagnetic Brillouin zone boundary perpendicular to $c$-axis. In other words, the conventional fully-gapped $s$-wave superconductivity is forbidden in this type of antiferromagnetic superconductors, irrelevant to the pairing mechanism, as long as the Fermi surface crosses a zone boundary.
UCoGe and UPd$_2$Al$_3$ are candidates for unconventional superconductors possessing hidden symmetry-protected line nodes, peculiar to non-symmorphic magnetic space groups.
\end{abstract}


\maketitle

In the research field of superconductivity, its coexistence of magnetism is a very interesting topic. Such coexistence between superconductivity and magnetism is often discovered in the U-based heavy-fermion superconductors. For example, UPd$_2$Al$_3$ shows an antiferromagnetic transition at $T_N=14$K, and then coexists with unconventional superconductivity below $T_c=2$K~\cite{Geibel,Krimmel}. UGe$_2$~\cite{Saxena}, URhGe~\cite{Aoki}, and UCoGe~\cite{Huy} encounter a superconducting transition in the ferromagnetic phase. A rare reentrant superconductivity has been discovered under the magnetic field~\cite{Levy}.
Theoretically, in UPd$_2$Al$_3$, it was discussed that a spin-singlet superconductivity with horizontal line nodes occurs via the virtual exchange of magnetic excitons~\cite{Sato,Miyake}. 
In the ferromagnetic superconductors, it has been considered that the Ising-like ferromagnetic fluctuation can lead to a spin-triplet pairing state~\cite{Hattori,Tada}, and many fascinating phenomena including the odd $H-T$ phase diagram have been studied~\cite{Mineev1,Mineev2,Mineev3,Tada2,HattoriK,Tada3}. 
However, in spite of the growing interest, the properties characteristic of the coexisting phase is less well understood systematically. In this situation, the group-theoreical classification, which provides definite statements independent of the details of materials, plays an important role.

It is well-known that the superconducting states are classified into the irreducible representations (IRs) under a given point group symmetry~\cite{Volovik1,Volovik2,Ueda}. Such classification provides useful information in analyzing the nodal structure of various unconventional superconductors~\cite{Sigrist}. 
Also, another development of gap classification based on the space group symmetry~\cite{Izyumov,Yarzhemsky1,Yarzhemsky2,Micklitz} gives us the correct way to take into account small representations at Brillouin zone (BZ) boundary in non-symmorphic space groups~\cite{Bradley, Bradley2}.
T. Micklitz and M. R. Norman~\cite{Micklitz} demonstrated in the pioneering work that new types of symmetry-protected nodes can appear at the BZ boundary.
As for the coexisting phase, there is little progress of gap classification considering the ordered moment~\cite{Mineev4,Samokhin}.

In this Letter, we extend the gap classification into {\it non-symmorphic magnetic} space groups~\cite{Bradley, Bradley2}. The results can be applied to the analysis of superconductivity in UCoGe or UPd$_2$Al$_3$. In these compounds, the magnetic ordered phase belongs to the non-symmorphic magnetic (type III or IV Shubnikov) space groups due to the presence of time-reversal symmetry with a point group operation and/or a non-primitive translation. Here, we show the following nontrivial consequences within the weak-coupling BCS theory.
The UCoGe-type ferromagnetic superconductors have horizontal line nodes (gap zeros) on either $k_z=0$ or $k_z=\pm\pi/c$ plane. In UPd$_2$Al$_3$-type antiferromagnetic superconductors, $A_g$ gap functions always have line nodes on $k_z=\pm\pi/c$ plane ({\it i.e.} the magnetic BZ face), in other words, the conventional fully-gapped $s$-wave superconductivity is forbidden. (Needless to say, in both cases, it is necessary that the Fermi surface crosses nodal planes.) Thus, UCoGe and UPd$_2$Al$_3$ are a candidate of unconventional superconductors possessing hidden symmetry-protected line nodes, peculiar to non-symmorphic magnetic space groups.

\begin{figure}[t]
\centering
\includegraphics[width=8cm,clip]{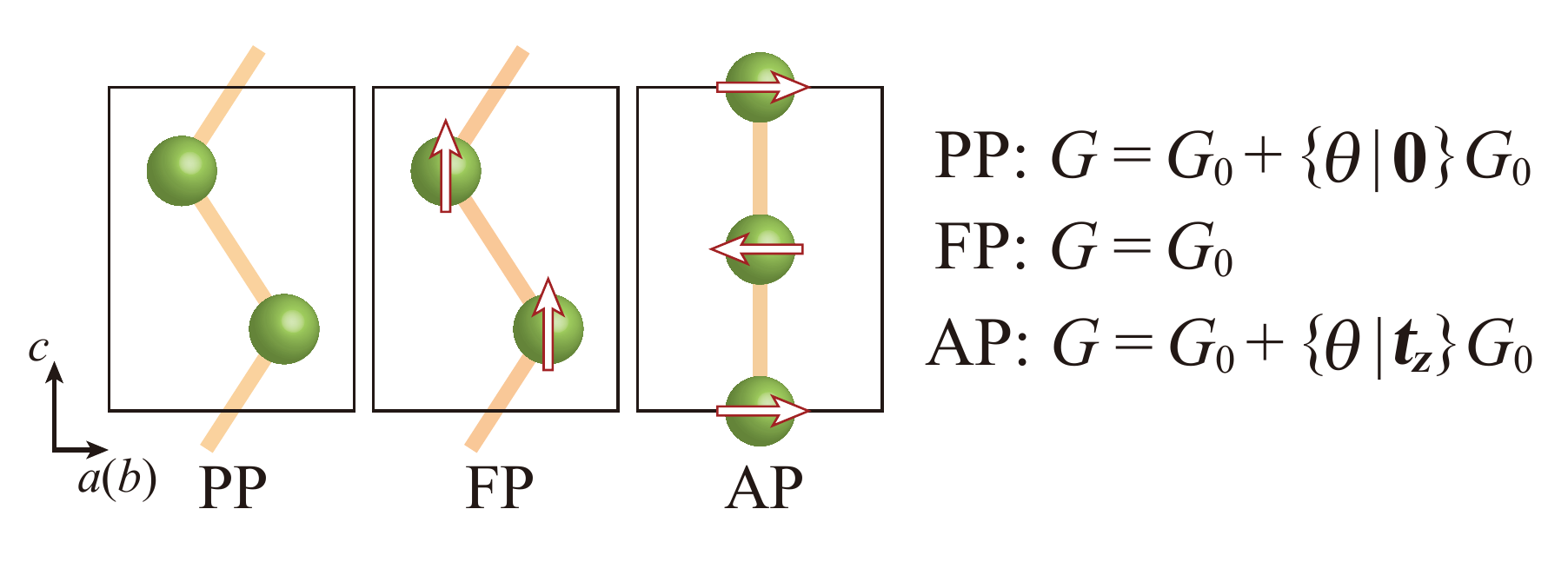}
\caption{(Color online) Three types of space groups considered in this letter. $\theta$ denotes a time-reversal operation. Left panel is a side view of the unit cell in the typical structures. PP and FP contain a zigzag structure, and AP possesses staggered ordered moments along the $c$-axis,  in which the orientations are in the $ab$-plane.}
\label{fig:0}
\end{figure}

{\it Set up ---}
First, we focus on a space group $G_0$ that is given as a coset decomposition $G_0=\{E|\bm 0\}T+\{2_z|\bt_z\}T+\{I|\bm 0\}T+\{\sigma_h|\bt_z\}T$, where the translation group $T$ defines a Bravais Lattice, and $\bt_z=\frac{c}{2}{\bm e}_c$ is a non-primitive translation along the $c$-axis. The notation $\{p|{\bm a}\}$ is a conventional Seitz space group symbol with a point-group operation $p$ and a translation $\bm a$. 
$E$ denotes an identity operation, $2_z$ a $\pi$-rotation around $c$-axis, $I$ a spatial-inversion, and $\sigma_h$ a mirror about $ab$-plane.
In this letter, we consider the three types of systems given in Fig.~\ref{fig:0}. PP, FP, and AP correspond to paramagnetic, ferromagnetic, and antiferromagnetic phase, respectively. Unless otherwise assigned, the spin-orbit coupling is included in all systems. Note that the space group $G$ of PP is the same as discussed in Ref.~\cite{Micklitz}.

{\it Method ---}
Let $\gamma_\bk$ be a small representation~\cite{note1} of a little group $K_\bk$, which represents the Bloch state with the crystal momentum $\bk$. We should note that the (zero-momentum) Cooper pairs have to be formed between the degenerate states present at $\bk$ and $-\bk$ within the BCS theory. Therefore, these two states should be connected by some symmetry operations except for an accidentally degenerate case. As a result, the representation of Cooper pair wave functions $P_\bk$ can be constructed from $\gamma_\bk$ as summarized in Refs.~\cite{Yarzhemsky1,Yarzhemsky2,Micklitz}. Here, we do not repeat the details of the prescription, and instead, indicate the practical procedure step by step.

{\it Results ---}
Through the present letter, we only consider the Cooper pairs in the basal plane ($k_z= 0$) and the zone face ($k_z=\pm\pi/c$). In both planes, the little groups $K_\bk$ are given by the following coset decompositions,
\begin{align}
K_\bk\!=\!\left\{\begin{aligned}
&\{E|{\bm 0}\}T\!+\!\{\sigma_h|{\bt_z}\}T\!+\!\{\theta I|{\bm 0}\}T\!+\!\{\theta 2_z|\bt_z\}T &&\!\!\!\!:{\rm PP}\\
&\{E|{\bm 0}\}T\!+\!\{\sigma_h|{\bt_z}\}T &&\!\!\!\!:{\rm FP} \\
&\{E|{\bm 0}\}T\!+\!\{\sigma_h|{\bt_z}\}T\!+\!\{\theta I|\bt_z\}T\!+\!\{\theta 2_z|{\bm 0}\}T &&\!\!\!\!:{\rm AP}
\end{aligned}
\right.\label{eq:2}
\end{align}
To obtain the small representations, it is sufficient to see the (projective) IRs of the corresponding little co-groups $\bar{K}_\bk=K_\bk/T$ with the appropriate factor systems~\cite{Bradley,Bradley2}. We denote them by $\bar{\gamma}_\bk$. Here, we specifies the elements of $\bar{K}_\bk$ as the representatives $r=\{p|{\bm a}\}$ of the decompositions \eqref{eq:2}. $\bar{\gamma}_\bk$ can be obtained by calculating the IRs for the unitary part of $\bar{K}_\bk$, and then inducing them by an anti-unitary operation~\cite{Murthy}. In Table\;\ref{table1}, we summarize the characters of $\bar{\gamma}_\bk$ for the unitary operations in $\bar{K}_\bk$. Now, the corresponding small representations are given by $\gamma_\bk(g)=\bar{\gamma}_\bk(r)F_\bk(t)$ where $g=rt$ for $g\in K_\bk$ and $t\in T$. $F_\bk$ is the IR of $T$ defined by $F_\bk(t)=e^{-i\bk\cdot\bt}$ for $t=\{E|\bt\}$. From Table\;\ref{table1}, we can see that the IRs of $\bar{K}_\bk$ in PP and AP become two-dimensional, which reflect the Kramers degeneracy for the anti-unitary operations $\{\theta I|{\bm 0}\}$ and $\{\theta I|{\bt_z}\}$. In the case of AP, since the non-primitive translation included in $\{\theta I|\bt_z\}$ cancels out the phase factor arising from that in $\{\sigma_h|\bt_z\}$, $\bar{\gamma}_\bk(\{\sigma_h|{\bm t}_z\})$ in the zone face is the same in the basal plane.
This situation is in sharp contrast to the case of PP.

\begin{table}[t]
\caption{The characters of $\bar{\gamma}_\bk$ in the case of PP, FP, and AP. Upper and lower expressions in PP and FP correspond to the two non-equivalent IRs.}\label{table1}
\begin{tabular}{>{\centering\arraybackslash}p{0.5cm}>{\centering\arraybackslash}p{1.1cm}>{\centering\arraybackslash}p{1.1cm}}
\multicolumn{3}{c}{Basal plane} \\ \hline \hline
$\bar{K}_\bk$      & $\{E|{\bm 0}\}$ & $\{\sigma_h|{\bm t}_z\}$   \\ \hline 
PP & $2$         & $0$  \\
FP & $1$         & $\pm i$                \\
AP & $2$         & $0$               \\ \hline
\end{tabular}
\hspace{0.3cm}
\begin{tabular}{>{\centering\arraybackslash}p{5mm}>{\centering\arraybackslash}p{1.1cm}>{\centering\arraybackslash}p{1.1cm}}
\multicolumn{3}{c}{Zone face} \\ \hline \hline
$\bar{K}_\bk$  & $\{E|{\bm 0}\}$ & $\{\sigma_h|{\bm t}_z\}$   \\ \hline 
PP & $2$         & $\pm2i$  \\
FP & $1$         & $\pm i$ \\
AP & $2$         & $0$       \\ \hline
\end{tabular}
\end{table}

\begin{table}[t]
\caption{The characters of $\bar{P}_\bk$ in PP, FP, and AP. }
\begin{tabular}{>{\centering\arraybackslash}p{1cm}>{\centering\arraybackslash}p{1.3cm}>{\centering\arraybackslash}p{1.3cm}>{\centering\arraybackslash}p{1.3cm}>{\centering\arraybackslash}p{1.3cm}}
\multicolumn{5}{c}{Basal plane} \\ \hline \hline
$M_\bk/T$  & $\{E|{\bm 0}\}$ & $\{2_z|{\bm t}_z\}$ & $\{I|{\bm 0}\}$ & $\{\sigma_h|{\bm t}_z\}$  \\ \hline
PP & $4$ & $2$ & $-2$ & $0$ \\
FP & $1$ & $1$ & $-1$ & $-1$                \\
AP & $4$ & $2$ & $-2$ & $0$               \\ \hline \\
\multicolumn{5}{c}{Zone face} \\ \hline \hline
$M_\bk/T$  &  $\{E|{\bm 0}\}$ & $\{2_z|{\bm t}_z\}$ & $\{I|{\bm 0}\}$ & $\{\sigma_h|{\bm t}_z\}$  \\ \hline
PP & $4$ & $-2$ & $-2$ & $4$ \\
FP & $1$ & $-1$ & $-1$ & $1$         \\
AP & $4$ & $-2$ & $-2$ & $0$               \\ \hline
\end{tabular}\label{table2}
\end{table}

Next, we consider the representation of the Cooper pairs $P_\bk$. 
In the space group operation $d$ connecting two states of the paired electrons, its rotation/inversion part $p_d$ meets $p_d\bk=-\bk$ modulo a reciprocal lattice vector. In the present cases, $\{I|{\bm 0}\}$ and $\{2_z|{\bm t}_z\}$ are the candidates for the operator $d$ in FP, while $\{\theta|{\bm 0}\}$ $(\{\theta|{\bm t}_z\})$ is also in PP (AP). Regardless of the choice of $d$, $M_\bk=K_\bk+dK_\bk$ is identical to the space group $G$. Taking into account the antisymmetry of the Cooper pairs and the degeneracy of the two states, we can regard $P_\bk$ as an antisymmetrized Kronecker square~\cite{Bradley}, with zero total momentum, of the induced representation $\gamma_\bk\uparrow M_\bk$. In the systematic way, this is obtained by using the double coset decomposition and the corresponding Mackey-Bradley theorem~\cite{Mackey,Bradley3,Bradley}. The obtained results are summarized in Table\;\ref{table2}. Here, $\bar{P}_\bk$ is the representation of $M_\bk/T$ to meet $P_\bk(g)=\bar{P}_\bk(r)$ where $g=rt$ for $g\in M_\bk$, $r\in M_\bk/T$, and $t\in T$. In the case of AP, the character of $\bar{P}_\bk$ for $\{\sigma_h|{\bm t}_z\}$ is equal to zero even in the zone face, different from the case of PP. This comes from the difference of $\bar{\gamma}_\bk(\{\sigma_h|{\bm t}_z\})$ in Table\;\ref{table1}. On the other hand, since $\bar{P}_\bk$ in FP is one-dimensional representation, only one IR is allowed both in the basal plane and the zone face.

Finally, we reduce the representation $\bar{P}_\bk$ into the IRs. In both planes, there are four IRs, $A_g,B_g,A_u$, and $B_u$ since the coset group $M_\bk/T$ is isomorphic to the point group $C_{2h}$~\cite{note2}. The results are summarized in Table\;\ref{table3}. Note that the gap functions should be zero, which means the appearance of gap nodes, if the corresponding IRs do not exist in the reduction of $\bar{P}_\bk$~\cite{Izyumov,Yarzhemsky1,Yarzhemsky2}.
The absence of $A_u$ in PP corresponds to the emergent horizontal line node of the recently proposed $E_{2u}$ state in UPt$_3$ superconductors~\cite{Micklitz,Nomoto,Yanase3,Kobayashi1}. On the other hand, in the case of AP, we find that $A_g$ does not appear in the zone face, in other words, $A_g$ possesses horizontal line nodes in the zone face ({\it i.e.} the magnetic BZ boundary). This means that the conventional fully-gapped $s$-wave superconductivity is forbidden, if the Fermi surfaces cross $k_z=\pm\pi/c$ plane. 
Emergence of line nodes in the zone boundary in antiferromagnetic superconductors was studied based on the microscopic theory in Refs.~\cite{Fujimoto,Yanase}.
This peculiar example can be realized in the superconductivity in UPd$_2$Al$_3$ as discussed below. In the case of FP, only odd-parity pairing is allowed in both planes, due to the absence of Kramers degeneracy. 
$A_u$ is forbidden in the zone face, and $B_u$ is forbidden in the basal plane. Therefore, the line nodes always appear, as for as the Fermi surface crosses $k_z=0$ and $k_z=\pm\pi/c$ planes. It should be noted that the emergence of such nodal structure does not depend on the pairing mechanism.
These are the main results of this Letter. Note that these results are applicable to not only conventional magnetic-dipole ordered states, but also magnetic multipole ordered states.

\begin{table}[t]
\caption{The reduction of $\bar{P}(\bk)$ to the IRs of $C_{2h}$ in the case of PP, FP, and AP. }
\begin{tabular}{>{\centering\arraybackslash}p{0.8cm}>{\centering\arraybackslash}p{2.4cm}}
\multicolumn{2}{c}{Basal plane} \\ \hline \hline
PP & $A_g+2A_u+B_u$ \\
FP & $A_u$                \\
AP & $A_g+2A_u+B_u$  \\ \hline 
\end{tabular}
\hspace{0.3cm}
\begin{tabular}{>{\centering\arraybackslash}p{0.8cm}>{\centering\arraybackslash}p{2.4cm}}
\multicolumn{2}{c}{Zone face} \\ \hline \hline
PP & $A_g+3B_u$     \\
FP & $B_u$          \\
AP & $B_g+A_u+2B_u$ \\ \hline
\end{tabular}\label{table3}
\end{table}

{\it Discussion ---}
Now we discuss several U-based materials in non-symmorphic magnetic space groups. First, we focus on the case of AP. The space group $G=G_0+\{\theta|\bt_z\}G_0$ corresponds to $P_b 2_1/m$ (the unique axis is chosen to be $c$-axis). Its typical example is the antiferromagnetic phase of UPd$_2$Al$_3$, in which the ordering vector is ${\bm Q}=(0,0,\pi/c)$ and the orientations of moments are in the basal plane.
Many experimental observations~\cite{Sato,Kyogaku,Bernhoeft,Watanabe,Shimizu} imply the presence of horizontal line nodes at the magnetic BZ boundary perpendicular to the $c$-axis. Moreover, it has been expected that the gap function belongs to $A_{g}$ IR~\cite{Miyake,Kyogaku2,Feyerherm,Matsuda} and the Fermi surfaces cross $k_z=\pm\pi/c$ plane~\cite{Inada,Inada2,Sandratskii,Inada3}. 
Following our results, the expected horizontal line nodes are symmetry-protected nodes in the non-symmorphic magnetic space groups. 

Next, let us consider the case of FP. Its candidates are hotly-debated ferromagnetic superconductors, UCoGe,  URhGe and UGe$_2$.
In the paramagnetic phase, the space group of UCoGe and URhGe is $Pnma1'$, while UGe$_2$ possesses symmorphic $Cmmm1'$. In the ferromagnetic phase, the former two belong to FP as shown below, while the latter does not meet the condition.
In the former two, since the ordered moments align parallel to $c$-axis in the ferromagnetic phase, the space group is reduced from $Pnma1'$ into $Pn'm'a$. This group is given explicitly by $G=G_0'+\{\theta 2_y|\bt_y\}G_0'$ where $G_0'=\{E|{\bm 0}\}T+\{2_z|{\bm t}_n\}T+\{I|{\bm 0}\}T+\{\sigma_h|{\bm t}_n\}T$ with $\bt_y=\frac{b}{2}{\bm e}_y$ and $\bt_n=\frac{a}{2}{\bm e}_x+\frac{c}{2}{\bm e}_z$. Here, $a,b$, and $c$ are the lattice parameters and $2_y$ represents the $\pi$-rotation around $b$-axis. At first glance, this seems rather different from that of FP. However, considering a general point in the basal plane and the zone face, we can ignore the anti-unitary part because this is not the element of the little groups. Therefore, the only difference is that the non-primitive translation becomes $\bt_n$ instead of $\bt_z$. As a result, $\bar{\gamma}_\bk(\{\sigma_h|\bt_z\})$ in the zone face changes from $\bar{\gamma}_\bk(\{\sigma_h|\bt_z\})=\pm i$ to $\pm i e^{-ik_x/2}$. We can easily confirm that this change does not affect the final results given in Table\;\ref{table2} and \ref{table3}. Therefore, any superconductivity in these materials have line nodes in either the basal plane or the zone face, at least, in the weak coupling limit. 

Now, let us consider the nodal structure of superconductivity in UCoGe in details. Unfortunately, the Fermi surfaces of this compound have not been established in experiments~\cite{Czekala,Aoki2,Fujimori}, however, the first-principles calculations show the existence of many complicated Fermi surfaces, some of which cross the $k_z=0$ and $k_z=\pm\pi/c$ planes~\cite{Czekala}. Therefore, our results suggest the existence of horizontal line nodes in the coexistent phase of UCoGe. Both power law behaviors of the spin-lattice relaxation rate $1/T_1\sim T^3$~\cite{Ohta,Ohta2} and the thermal conductivity $\kappa_S/\kappa_N\sim T^2$~\cite{Howald} are consistent with our prediction. Note that in the pressure-temperature phase diagram, the ferromagnetic transition seems to have little affect on the superconductivity~\cite{Hassinger,Hassinger2}. 
In the absence of magnetism, the straightforward calculation shows that the space group $Pnma1'$ leads to the same nodal structure as the case of PP in Table\;\ref{table3}, by regarding the IRs of $D_{2h}$ as those of $C_{2h}$ with the compatibility relation. Therefore, the line nodes in the basal plane of $B_u$ IR ($B_{2u}$ and $B_{3u}$ IRs of $D_{2h}$ group) are forbidden in the paramagnetic phase, which is consistent with the Blount's theorem of the triplet superconductors~\cite{Blount}. Therefore, we may expect that the realized gap function belongs to $A_u$ IR ($A_{1u}$ or $B_{1u}$ in the paramagnetic phase) and has the line nodes at $k_z=\pm\pi/c$ plane. As for the $A_u$ gap functions in the coexistent phase, there should be an additional point node at $k_x=k_y=0$, which is regarded as Weyl nodes~\cite{Mineev4}. Thus, the hybrid gap structure of line and point nodes would be realized such as proposed in URu$_2$Si$_2$~\cite{Shibauchi} and UPt$_3$~\cite{Joynt}. 

\begin{figure}[t]
\centering
\includegraphics[width=8cm,clip]{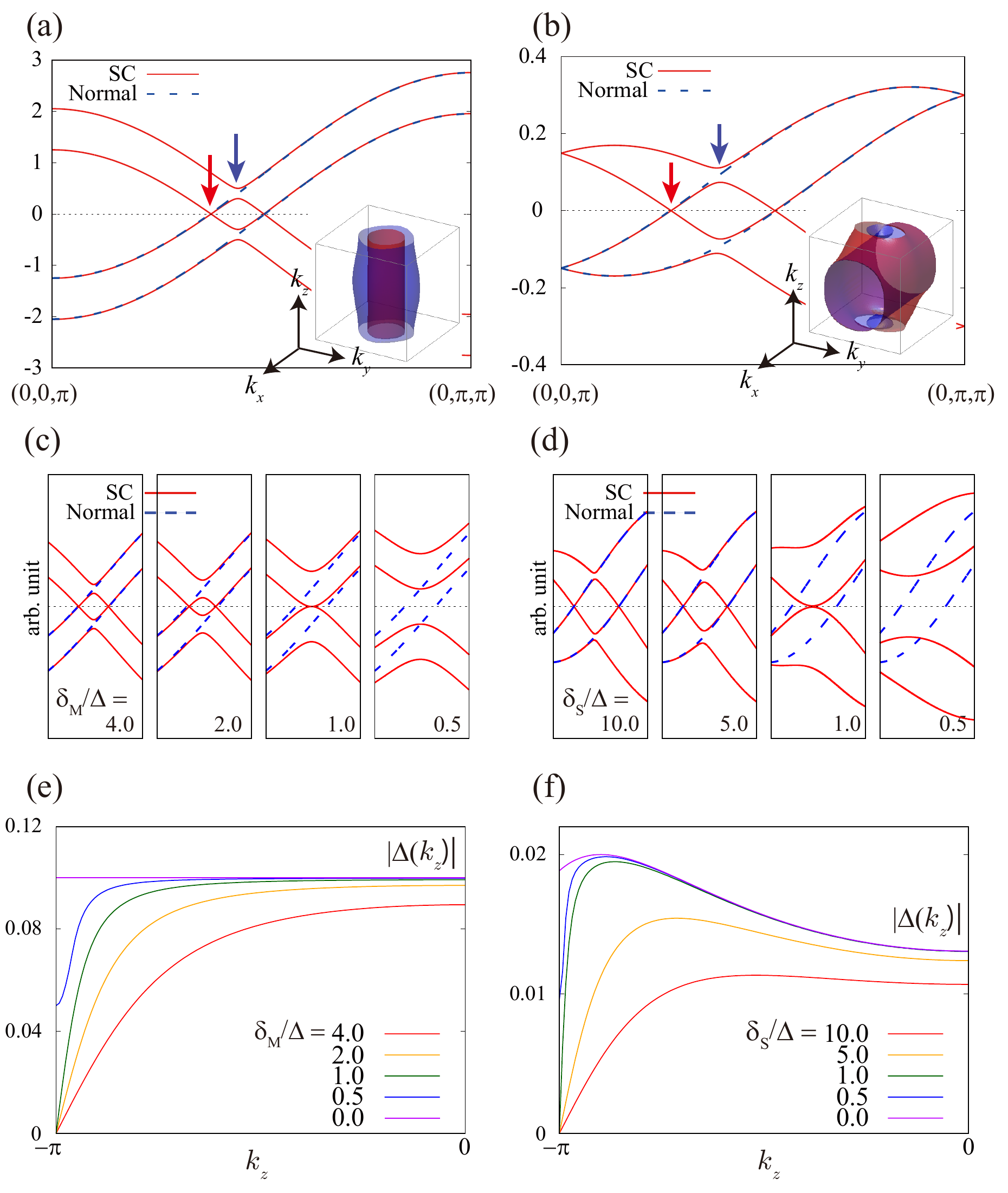}
\caption{(Color online) Typical band structure of the BdG Hamiltonian assuming (a) the $A_g$ gap function in the AP model of UPd$_2$Al$_3$ and (b) the $A_u$ gap function in the FP model of UCoGe~\cite{model}. The lattice constants are set to be unity. Measure of energy is unit of the nearest neighbor hopping integral. Blue dashed lines correspond to the original band in the normal state. The inset shows the corresponding Fermi surface. (c) and (d) are the enlarged figure of the Bogoliubov band for several $\Delta$.
(e) and (f) show the $k_z$ dependence of the excitation gap amplitude $|\Delta(k_z)|$ on the Fermi surface at $k_x=0$ for several $\delta_M$ and $\delta_S$.}
\label{fig:1}
\end{figure}

Finally, we demonstrate the above-mentioned group theoretical arguments by using a specific model, and discuss the stability of horizontal line nodes on the zone face. We consider a Bogoliubov-de Gennes (BdG) Hamiltonian of an AP (FP) superconductivity with $A_g$ ($A_u$) gap structure as a minimal model of UPd$_2$Al$_3$ (UCoGe)~\cite{model}.
Fig.\;\ref{fig:1}(a) depicts the band structure along the high-symmetry line $(0,0,\pi)$-$(0,\pi,\pi)$ in the AP model of UPd$_2$Al$_3$, and its inset shows the Fermi surface.
Fig.\;\ref{fig:1}(b) is the case of UCoGe.
In both Figs.\;\ref{fig:1}(a) and (b), we can find that the superconducting gap remains closed at the Fermi level in the particle-hole symmetric Bogoliubov band (red arrows), while a gap is open at the inter-band crossing point far from the Fermi level (blue arrows). The emergence of gap zero on the zone face is fully consistent with the group theoretical arguments. On the other hand, the group theory does not say anything about the inter-band gap opening, since the above-mentioned arguments are based on the intra-band pairs. As readily understood, if the inter-band gap is sufficiently large, then the symmetry-protected intra-band gap nodes can be lost.
In our models, the emergence of gap nodes is controlled by three parameters, $\Delta$, $\delta_M$, and $\delta_S$, which correspond to respectively the gap amplitude, the strength of the magnetic order and the spin-orbit coupling~\cite{model}.
$\delta_M$ and/or $\delta_S$ lift the band degeneracy on the zone face~\cite{split}.
Figs.\;\ref{fig:1}(c) and (d) are the Bogoliubov band structure for several $\Delta$. As expected, we can find the vanishing of gap nodes for larger $\Delta$ via a kind of Lifshitz transition.
Figs.\;\ref{fig:1}(e) and (f) show $k_z$ dependence of the excitation gap amplitude $|\Delta(k_z)|$ on the Fermi surface at $k_x=0$ for several $\delta_M$ and $\delta_S$. 
For relatively large $\delta_M (\delta_S)$, $\Delta(k_z)$ behaves like $\Delta(k_z)\sim \cos (k_z/2) $. On the other hand, for relatively small $\delta_M (\delta_S)$, the gap amplitude sharply changes around the nodes on the zone face, which was also discussed in Ref.~\cite{Yanase}. In the limit of $\delta_M (\delta_S)=0$, the nodal structure is completely lost.

In our model, such $k_z$ dependence just comes from the unitary matrix diagonalizing the BdG Hamiltonian. It implies that even the BCS approximation of purely local (on-site) interactions, such as the conventional electron-phonon interactions, can induce the anisotropic gap structure in the non-symmorphic magnetic superconductors. 
In the realistic situations, the band splittings on the zone face will be sufficiently larger than the gap amplitude. Therefore, it is expected that the present nodal structure can be observed as the usual power-law behavior in thermodynamic and/or transport properties at low temperatures. 
Consequently, nontrivial symmetry-protected line nodes in the non-symmophic magnetic space groups will be observed in magnetic superconductors UCoGe or UPd$_2$Al$_3$.

\begin{acknowledgements}
We acknowledge K. Hattori, Y. Yanase, S. Fujimoto and K. Shiozaki for valuable discussions. This work was partly supported by JSPS KAKENHI Grant No.15H05745, 15H02014,  15J01476, 16H01081, and 16H04021.
\end{acknowledgements}

\pagebreak
\widetext
\begin{center}
\vspace{1cm}
\textbf{\large Supplemental Materials: \\ \vspace{3mm} Symmetry protected line nodes in non-symmorphic magnetic space groups: Applications to  UCoGe and UPd$_2$Al$_3$}
\end{center}
\setcounter{equation}{0}
\setcounter{figure}{0}
\setcounter{table}{0}
\setcounter{page}{1}
\makeatletter
\renewcommand{\theequation}{S\arabic{equation}}
\renewcommand{\thefigure}{S\arabic{figure}}
\renewcommand{\bibnumfmt}[1]{[S#1]}
\renewcommand{\citenumfont}[1]{S#1}
\newcommand{\bx}{{\bm x}}

\section{Model Hamiltonian}
In this supplementary material, we show the details of our model Hamiltonians, which mimic UPd$_2$Al$_3$ or UCoGe superconductors. For simplicity, we set the lattice constant to unity, and consider a single orbital on each U site.

\subsection{UPd$_2$Al$_3$}
The antiferromagnetic phase of UPd$_2$Al$_3$ belongs to the magnetic space group $P_b 2_1/m$ (the unique axis is chosen to be $c$-axis). In the unit cell, two U atoms are placed at ${\bm x}_1=(0,0,0)$ and ${\bm x}_2=(0,0,\frac{1}{2})$. A minimal tight-binding Hamiltonian contains four orbitals, corresponding to two atoms (sublattice) and the spin-1/2 degrees of freedom. The Hamiltonian $\mathcal{H}$ is defined by,
\begin{align}
\mathcal{H}=\sum_{\bk}\sum_{\alpha\beta}\sum_{\sigma\sigma'}h_{\alpha\sigma,\beta\sigma'}(\bk)c^\dagger_{\alpha\sigma}(\bk) c_{\beta\sigma'}(\bk), \label{eq1}
\end{align}
where $c^\dagger_{\alpha\sigma}(\bk)$ ($c_{\beta\sigma'}(\bk)$) is a creation (annihilation) operator of electrons with spin $\sigma\,(\sigma')=\uparrow,\,\downarrow$ on an atom $\alpha\,(\beta)=1,\,2$. Note that, through this paper, we have used the site dependent Fourier transformation defined as,
\begin{align}
c^\dagger_{\alpha\sigma}(\bk)&=\frac{1}{\sqrt{N}}\sum_{{\bm R}}e^{i\bk\cdot({\bm R}+\bx_\alpha)}c^\dagger_{\alpha\sigma}({\bm R}),
\end{align}
where $N$ is a total number of unit cell, ${\bm R}$ is a lattice vector, and ${\bm x}_\alpha$ is a relative position for the site $\alpha$ in the unit cell. In this case, $h_{\alpha\sigma,\beta\sigma'}(\bk)$ in the matrix form is given by,
\begin{align}
h(\bk)=\varepsilon_0(\bk)\tau^0\otimes\sigma^0+\varepsilon_1(\bk)\tau^x\otimes\sigma^0+\delta_M\tau^z\otimes\sigma^x, \label{eq:updal}
\end{align}
where $\varepsilon_0(\bk)=-2t_{xy} (\cos k_x+\cos k_y) -2t_z' \cos k_z-\mu$ and $\varepsilon_1(\bk)=-2 t_z \cos \frac{k_z}{2}$. $\tau^\mu$ and $\sigma^\mu$ ($\mu=0,x,y,$ and $z$) are the Pauli matrices acting on the sublattice and the spin degrees of freedom, respectively. The third term of Eq.\,\eqref{eq:updal} tunes the magnitude of the staggered magnetic moment along $a$-axis in the antiferromagnetic phase.

In the superconducting state, the anomalous part $\Psi^\varGamma$ given by 
\begin{align}
\Psi^\varGamma=\sum_{\bk}\sum_{\alpha\beta}\varphi^\varGamma_{\alpha\sigma,\beta\sigma'}(\bk)c^\dagger_{\alpha\sigma}(\bk) c_{\beta\sigma'}^\dagger(-\bk)
\end{align}
should be added in the Hamiltonian $\mathcal{H}$. Here, $\varphi^\varGamma_{\alpha\sigma,\beta\sigma'}(\bk)$ is the corresponding order parameter. The superscript $\varGamma$ denotes an IR of the point group $C_{2h}$. 
For the $A_{g}$ spin-singlet pairing state discussed in Fig.\,1, we can take momentum-independent (constant) order parameter,
\begin{align}
\varphi^{A_g}(\bk)=\Delta \tau^0\otimes (i\sigma^y),
\end{align}
where $\Delta$ is the gap amplitude. 

The band structure in Fig.\,1(a) is the result for the parameters $(t_{xy},t_z,t_z',\delta_M,\Delta,\mu)=(1.0,0.4,0.1,0.4,0.1,-2.0)$. Figs.\,1(c) and (e) were obtained by changing $\Delta$ and $\delta_M$, respectively.

\subsection{UCoGe}
The ferromagnetic phase of UCoGe belongs to the magnetic space group $Pn'm'a$. U atoms are placed at $\bx_1=(x,\frac{1}{4},z), \bx_2=(\frac{1}{2}-x,\frac{3}{4},z-\frac{1}{2}), \bx_3=(1-x,\frac{3}{4},1-z),$ and $\bx_4=(\frac{1}{2}+x,\frac{1}{4},\frac{3}{2}-z)$ in the unit cell, where $x=0.0101,z=0.7075$~\cite{Canepa}. The ferromagnetic moments are aligned along $c$-axis. In this case, $h(\bk)$ in Eq.\,\eqref{eq1} is given by,
\begin{align}
h(\bk)=h_0(\bk)+h_S(\bk)+h_M,
\end{align}
where $h_0(\bk)$, $h_S(\bk)$, and $h_M$ represent the hopping integral, the spin-orbit coupling, and the exchange interaction with the magnetic moments, respectively. These are given by,
\begin{subequations}
\begin{align}
h_0(\bk)&=\begin{pmatrix}
\varepsilon_0(\bk) & \varepsilon_{12}(\bk) & \varepsilon_{13,-}(\bk) & \varepsilon_{14}(\bk) \\
\varepsilon_{12}^*(\bk) & \varepsilon_0(\bk) & \varepsilon_{14}(\bk) & \varepsilon_{13,+}(\bk) \\
\varepsilon_{13,-}^*(\bk) & \varepsilon_{14}^*(\bk) & \varepsilon_0(\bk) & \varepsilon_{12}^*(\bk) \\
\varepsilon_{14}^*(\bk) & \varepsilon_{13,+}^*(\bk) & \varepsilon_{12}(\bk) & \varepsilon_0(\bk) \\
\end{pmatrix}\otimes \sigma^0,\\
h_S(\bk)&=\delta_S\, \sin k_y\, {\rm diag}(1,-1,-1,1)\otimes\sigma^z,\\
h_M &= \delta_M\,{\rm diag}(1,1,1,1)\otimes\sigma^z.
\end{align}
\end{subequations}
Here, we only consider a simple spin-orbit coupling term, which lifts the band degeneracy on the zone face. Each element of the hopping matrix $\varepsilon_0(\bk)$, $\varepsilon_{12}(\bk)$, $\varepsilon_{13,\pm}(\bk)$, and $\varepsilon_{14}(\bk)$ is given by,
\begin{subequations}
\begin{align}
\varepsilon_0(\bk)&=-2t_y\cos k_y-\mu\\
\varepsilon_{12}(\bk)&=4t_{12}\cos\frac{k_y}{2}\cos\frac{k_z}{2}(e^{-(4x-1)k_x/2i}+\lambda_1 e^{-(4x+1)k_x/2i}),\\
\varepsilon_{13,\pm}(\bk)&=2t_{13}\cos\frac{k_y}{2}(e^{-(2z-1)k_zi}+\lambda_2 e^{-(2z-2)k_zi}) e^{\pm 2 xk_x i}, \\
\varepsilon_{14}(\bk)&=2t_{14}\cos\frac{k_x}{2}e^{-(4z-3)k_z/2i}.
\end{align}
\end{subequations}

In the superconducting phase, we set a typical $A_{u}$-type gap function,
\begin{align}
\varphi^{A_u}(\bk)=\Delta\,{\rm diag}(1,1,1,1)\otimes(\sin k_x i\sigma^x\sigma^y+\sin k_y i\sigma^y\sigma^y), \label{eq2}
\end{align}
where $\Delta$ is the gap amplitude. Note that each element of the (magnetic) point group $P$ is given by $P=\{E, 2_z, \theta 2_y, \theta 2_x, I, \sigma_h,\theta I2_y,\theta I2_x\}$, whose IRs are summarized in TABLE \ref{table1}~\cite{Mineev4}. Since the unitary transformation to the basis functions can connect the representations with different $\omega$ in TABLE \ref{table1}, these representations are equivalent in the sense of corepresentations. 
Eq.\,\eqref{eq2} corresponds to the case of $\omega=1$.
Gap functions with $\omega\neq 1$ can be simply obtained by the transformation $\varphi^\varGamma(\bk)\mapsto \omega^{-\frac{1}{2}}\varphi^\varGamma(\bk)$ without any change of the nodal structure.

The band structure in Fig.\,1(b) is the result for the parameters,
\begin{align*}
(t_y,t_{12},t_{13},t_{14},\lambda_1,\lambda_2,\delta_M,\delta_S,\Delta,\mu)=(0.1,0.4,1.0,0.8,0.8,0.8,0.2,0.2,0.02,-1.9).
\end{align*}
Figs.\,1(d) and (f) were obtained by changing $\Delta$ and $\delta_S$, respectively.

\begin{table}[t]
\caption{The character table of point group $P$ in UCoGe. $\omega$ is an arbitrary phase factor.}
\begin{tabular}{>{\centering\arraybackslash}p{0.8cm}>{\centering\arraybackslash}p{1.5cm}>{\centering\arraybackslash}p{1.5cm}>{\centering\arraybackslash}p{1.5cm}>{\centering\arraybackslash}p{1.5cm}>{\centering\arraybackslash}p{1.5cm}>{\centering\arraybackslash}p{1.5cm}>{\centering\arraybackslash}p{1.5cm}>{\centering\arraybackslash}p{1.5cm}} \hline\hline
IRs  & $E$ & $2_z$ & $\theta 2_y$ & $\theta 2_x$ & $I$ & $\sigma_h$ & $\theta I 2_y$ & $\theta I 2_x$ \\ \hline
$A_{g}$ & $1$ & $1$ & $\omega$ & $\omega$ & $1$ & $1$ & $\omega$ & $\omega$ \\
$B_{g}$ & $1$ & $-1$ & $\omega$ & $-\omega$ & $1$ & $-1$ & $\omega$ & $-\omega$ \\
$A_{u}$ & $1$ & $1$ & $\omega$ & $\omega$ & $-1$ & $-1$ & $-\omega$ & $-\omega$ \\
$B_{u}$ & $1$ & $-1$ & $\omega$ & $-\omega$ & $-1$ & $1$ & $-\omega$ & $\omega$ \\ \hline
\end{tabular}\label{table1}
\end{table}


\begin{thebibliography}{99}
\bibitem{note1} In this paper, we use the word ``representation'' both as a representation of a unitary group and a corepresentation of a non-unitary group.
\bibitem{note2} Strictly, $M_\bk/T$ is isomorphic to the gray point group of $C_{2h}$ in PP and AP. However, the corepresentations are trivial because the anti-unitary operations do not cause the extra degeneracy.
\bibitem{split} Note that, in order to lift the band degeneracy on the zone face, we require both the spin-orbit coupling and the exchange interaction with the magnetic moments in UCoGe, while only latter is necessary in UPd$_2$Al$_3$. This comes from the following reasons. Without the spin-orbit coupling, the direction of the magnetic moments can be freely chosen, and then, the extended time-reversal symmetry $\{\theta 2_z|\bt_n\}$ is restored in UCoGe. As a result, there remains two-fold degeneracy on the zone face due to the corresponding Kramers theorem, even in the presence of ferromagnetism. Note also that, in UCoGe, the spin-orbit coupling vanishes along $k_y=0$ line on the zone face, which is also protected by the Kramers theorem for the anti-unitary operation $\{\theta 2_y \sigma_h|\bt_y-\bt_n\}$. This implies that the present line nodes also disappear along $k_y=0$ line, which can be directly confirmed by the present group-theoretical arguments, and become arc line nodes similar to ones discussed in Ref.~\cite{Yanase3}.
\end{thebibliography}
\end{document}